\documentclass[superscriptaddress,onecolumn,secnumarabic,
amssymb,amsmath,nobibnotes,aps,prd,showkeys,showpacs,nofootinbib]{revtex4}

\usepackage[latin1]{inputenc}
\usepackage{graphicx}
\usepackage[english]{babel}

\usepackage{amsmath}
\usepackage{amssymb}
\usepackage{amsfonts}
\usepackage{colordvi}
\usepackage{psfrag}
\usepackage{color}

\begin{document}
\def\cL{{\cal L}}
\def\be{\begin{equation}}
\def\ee{\end{equation}}
\def\bea{\begin{eqnarray}}
\def\eea{\end{eqnarray}}
\def\beq{\begin{eqnarray}}
\def\eeq{\end{eqnarray}}
\def\tr{{\rm tr}\, }
\def\nn{\nonumber \\}
\def\e{{\rm e}}


\def\bef{\begin{figure}}
\def\eef{\end{figure}}
\newcommand{\ans}{ansatz }
\newcommand{\eeqn}{\end{eqnarray}}
\newcommand{\bd}{\begin{displaymath}}
\newcommand{\ed}{\end{displaymath}}
\newcommand{\mat}[4]{\left(\begin{array}{cc}{#1}&{#2}\\{#3}&{#4}
\end{array}\right)}
\newcommand{\matr}[9]{\left(\begin{array}{ccc}{#1}&{#2}&{#3}\\
{#4}&{#5}&{#6}\\{#7}&{#8}&{#9}\end{array}\right)}
\newcommand{\matrr}[6]{\left(\begin{array}{cc}{#1}&{#2}\\
{#3}&{#4}\\{#5}&{#6}\end{array}\right)}
\newcommand{\cvb}[3]{#1^{#2}_{#3}}
\def\lsim{\raise0.3ex\hbox{$\;<$\kern-0.75em\raise-1.1ex
e\hbox{$\sim\;$}}}
\def\gsim{\raise0.3ex\hbox{$\;>$\kern-0.75em\raise-1.1ex
\hbox{$\sim\;$}}}
\def\abs#1{\left| #1\right|}
\def\simlt{\mathrel{\lower2.5pt\vbox{\lineskip=0pt\baselineskip=0pt
           \hbox{$<$}\hbox{$\sim$}}}}
\def\simgt{\mathrel{\lower2.5pt\vbox{\lineskip=0pt\baselineskip=0pt
           \hbox{$>$}\hbox{$\sim$}}}}
\def\unity{{\hbox{1\kern-.8mm l}}}
\newcommand{\eps}{\varepsilon}
\def\ep{\epsilon}
\def\ga{\gamma}
\def\Ga{\Gamma}
\def\om{\omega}
\def\omp{{\omega^\prime}}
\def\Om{\Omega}
\def\la{\lambda}
\def\La{\Lambda}
\def\al{\alpha}
\newcommand{\ov}{\overline}
\renewcommand{\to}{\rightarrow}
\renewcommand{\vec}[1]{\mathbf{#1}}
\newcommand{\vect}[1]{\mbox{\boldmath$#1$}}
\def\tm{{\widetilde{m}}}
\def\mcirc{{\stackrel{o}{m}}}
\newcommand{\Dm}{\Delta m}
\newcommand{\dm}{\varepsilon}
\newcommand{\tanb}{\tan\beta}
\newcommand{\nbar}{\tilde{n}}
\newcommand\PM[1]{\begin{pmatrix}#1\end{pmatrix}}
\newcommand{\up}{\uparrow}
\newcommand{\down}{\downarrow}
\def\omE{\omega_{\rm Ter}}
%

\newcommand{\Dsusy}{{susy \hspace{-9.4pt} \slash}\;}
\newcommand{\DCP}{{CP \hspace{-7.4pt} \slash}\;}
\newcommand{\mc}{\mathcal}
\newcommand{\gr}{\mathbf}
\renewcommand{\to}{\rightarrow}
\newcommand{\gtc}{\mathfrak}
\newcommand{\wh}{\widehat}
\newcommand{\br}{\langle}
\newcommand{\kt}{\rangle}


\def\lsim{\mathrel{\mathop  {\hbox{\lower0.5ex\hbox{$\sim$}
\kern-0.8em\lower-0.7ex\hbox{$<$}}}}}
\def\gsim{\mathrel{\mathop  {\hbox{\lower0.5ex\hbox{$\sim$}
\kern-0.8em\lower-0.7ex\hbox{$>$}}}}}

\def\nn{\\  \nonumber}
\def\de{\partial}
\def\brf{{\mathbf f}}
\def\bbf{\bar{\bf f}}
\def\bF{{\bf F}}
\def\bbF{\bar{\bf F}}
\def\bA{{\mathbf A}}
\def\bB{{\mathbf B}}
\def\bG{{\mathbf G}}
\def\bI{{\mathbf I}}
\def\bM{{\mathbf M}}
\def\bY{{\mathbf Y}}
\def\bX{{\mathbf X}}
\def\bS{{\mathbf S}}
\def\bb{{\mathbf b}}
\def\bh{{\mathbf h}}
\def\bg{{\mathbf g}}
\def\bla{{\mathbf \la}}
\def\bmu{\mathbf m }
\def\by{{\mathbf y}}
\def\bmu{\mbox{\boldmath $\mu$} }
\def\bsig{\mbox{\boldmath $\sigma$} }
\def\bunity{{\mathbf 1}}
\def\cA{{\cal A}}
\def\cB{{\cal B}}
\def\cC{{\cal C}}
\def\cD{{\cal D}}
\def\cF{{\cal F}}
\def\cG{{\cal G}}
\def\cH{{\cal H}}
\def\cI{{\cal I}}
\def\cL{{\cal L}}
\def\cN{{\cal N}}
\def\cM{{\cal M}}
\def\cO{{\cal O}}
\def\cR{{\cal R}}
\def\cS{{\cal S}}
\def\cT{{\cal T}}
\def\eV{{\rm eV}}

\title{Evaporation/Antievaporation and energy conditions in alternative gravity}

\author{Andrea Addazi}

\affiliation{ Center for Field Theory and Particle Physics \& Department of Physics, Fudan University, 200433 Shanghai, China}

\date{\today}

\begin{abstract}
We discuss the evaporation and antievaporation instabilities of Nariai solution in extended theories of gravity.
These phenomena were explicitly shown in several different extensions of General Relativity,
suggesting that a universal cause is behind them. 
We show that evaporation and antievaporation are originated from deformations of energy conditions on the 
Nariai horizon. Energy conditions get new contributions from the extra propagating degrees of freedom, 
which can provide extra focalizing or antifocalizing terms in the Raychanduri equation. 
We also show two explicit examples in $f(R)$-gravity and Gauss-Bonnet gravity.


\end{abstract}

\maketitle

\section{Introduction}

Evaporation/antievaporation instabilities were firstly discovered by Bousso and Hawking in degenerate Schwarzschild-De Sitter BH (Nariai solutions), in context of quantum gravity coupled with a dilaton field \cite{Bousso:1997wi} -- elaborated later on in Refs. \cite{Nojiri:1998ph,Nojiri:1998ue,Elizalde:1999dw}. Intriguingly, classical evaporation/antievaporation were re-discovered in various different extensions of
General Relativity: f(R)-gravity, f(T)-gravity, Gauss-Bonnet gravity, Mimetic-gravity, string-inspired gravity, Bigravity, Bardeen gravity
and so on \cite{Nojiri:2013su,Nojiri:2014jqa,Sebastiani:2013fsa,Houndjo:2013qna,Oikonomou:2016fxb,Oikonomou:2015lgy,Addazi:2016hip,Katsuragawa:2014hda,Addazi:2017puj,Singh:2017qur}.
Recently, we have shown how that Evaporation/Antievaporation effects 
cause back-reactions which turns of the Bekenstein-Hawking radiation
\cite{Addazi:2016prb,Addazi:2017lat}.
However, the origin of this phenomena was not clarified in literature. 

\vspace{0.2cm}
{\it Why were evaporation/antievaporation instabilities 
found in 
so different extended theories of gravity?

Is there any universal cause behind evaporation/antievaporation instabilities? }
\vspace{0.2cm}

In this paper, we show how evaporation/antievaporation are related to the deformation of energy conditions in extended theories of gravity.
 In particular, the presence of extra degrees of freedom propagating in extended
theories of gravity can provide contributions to the energy-momentum tensor, i.e. they compose a {\it geometric
fluid} pervading the space-time \cite{Capozziello:2013vna,Capozziello:2014bqa,Addazi:2017rkc}.
In other words, the evaporation/antievaporation can be 
interpreted as outfalling/infalling fluxes of the geometric fluid inside the horizon. This is exactly the source of evaporation/antievaporation if and only if the extra terms provide
extra repulsion/attraction contributions to the Ricci tensor. In fact, new terms antifocalize/focalize space-time Cauchy hyper-surfaces so that
they are dynamically attracted outside/inside the BH interior.

The paper is organized as follows: in Section 2, we 
review general aspects of energy conditions in Extended theories of gravity,
in Section 3 we show our main arguments on evaporation and antievaporation, 
in Section 4 we shown our conclusions.

\section{Energy conditions in Extended theories of gravity}

Eulero-Lagrange equations of motion of locally Lorentz invariant Extended theories of gravity
can be written in a universal form as follows:
\be \label{EoM}
\mathcal{G}_{1}(\mathcal{I}^{i})(G_{\mu\nu}+H_{\mu\nu})=-8\pi G_{N}\mathcal{G}_{2}(\mathcal{I}^{i})T_{\mu\nu}\, ,
\ee
where $\mathcal{G}_{1,2}(\mathcal{I}^{i})$ are functionals of either curvature invariants
while $H_{\mu\nu}$ is a new geometrical tensor.
$\mathcal{G}_{1,2}(\mathcal{I}^{i})$ will modify matter gravity couplings.
For simplicity, we will consider the case $\mathcal{G}_{2}=1$.
$H_{\mu\nu}$ will correct the energy-momentum tensor with 
a new geometric term.
In fact, Eq.(\ref{EoM}) with $\mathcal{G}_{2}=1$ can be rewritten as 
\be \label{EoM2}
G_{\mu\nu}=-8\pi G_{N}\tilde{T}_{\mu\nu}\, ,
\ee
where 
\be \label{EoM3}
\tilde{T}_{\mu\nu}= \mathcal{G}_{1}^{-1}T^{\mu\nu}+\frac{1}{8\pi G_{N}}H_{\mu\nu}\, .
\ee

The general Raychaudhuri equation is 
\be \label{Ray}
\dot{\theta}+\frac{\theta^{2}}{3}+2(\sigma^{2}-\omega^{2})-W_{;\mu}^{\,i}=-R_{\mu\nu}W^{\mu}W^{\nu}\, ,
\ee
where $\sigma_{\mu\nu},\omega_{\mu\nu}$ are the shear and vorticity tensors;
$W^{\mu}$ is the the vector orthogonal to a D-dimensional Chauchy surface
satisfying  $W^{\mu}h_{\mu\nu}=0$, where $h_{\mu\nu}$ is the metric induced on he Chauchy manifold;
$\theta$ is the expansion scalar function, $\dot{\theta}=\frac{d\theta}{d\lambda}$ where $\lambda$ is an affine variable.
The optical Raychauduri equation has a zero vorticity tensor:
 \be \label{Ray2}
\dot{\theta}+\frac{\theta^{2}}{3}+2\sigma^{2}=-R_{\mu\nu}W^{\mu}W^{\nu}\, .
\ee

For completeness we also remind the formal definition of $\sigma,\omega,\dot{W},\theta$
in terms of the metric tensor and the vector $W$:
\be \label{def1}
h_{\mu\nu}=g_{\mu\nu}+W_{\mu}W_{\nu}\, ,
\ee
\be \label{def2}
\theta=h_{\mu}^{\nu}\nabla_{\nu} W_{\rho} h^{\mu\rho}\, ,
\ee
\be \label{def3}
\dot{W}^{\mu}=W_{\nu} \nabla^{\nu} W^{\mu}\, ,
\ee
\be \label{def4}
\sigma_{\mu\nu}=h_{(\mu}^{\rho} \nabla_{\rho} W_{\sigma} h^{\sigma}_{\nu)}-\frac{1}{3}h_{\mu\nu}h^{\delta}_{\rho}\nabla_{\delta} W_{\gamma} h^{\rho \gamma}\, ,
\ee
\be \label{def5}
\omega_{\mu\nu}=h_{[\mu}^{\delta} \nabla_{\delta} W_{\sigma} h^{\sigma}_{\nu]}\, . 
\ee
The Raychaudhuri equation is governed by GR field equation and 
it is constrained by energy conditions. For the completeness of our following discussions
we will briefly review the energy conditions followed by ordinary matter 
in GR. 

\vspace{0.3cm}

{\bf The Strong energy condition} states that
\be \label{TWW}
\tilde{T}_{\mu\nu}W^{\mu}W^{\nu}-\frac{1}{2}\tilde{T}W^{\mu}W_{\mu}\geq 0\, ,
\ee
where $W^{\mu}$ is a time-like 4-vector, i.e. $W^2=-1$.
This condition is related to an attractive nature of gravity, i.e. $R_{\mu\nu}W^{\mu}W^{\nu}\geq 0$. 

\vspace{0.3cm}

{\bf The Null energy condition} is 
\be \label{Tkk}
\tilde{T}_{\mu\nu}k^{\mu}k^{\nu} \geq 0\, , 
\ee
where $k^{\mu}$ is the null-like vector satisfying $k^{2}=0$.
The Strong energy condition inevitably implies that the Hamiltonian 
of systems is unbounded from below
and that $R_{\mu\nu}k^{\mu}k^{\nu}\geq 0$.
The null energy condition in Eq.(\ref{Ray2}) implies the focusing attractive 
nature of gravity. 

\vspace{0.3cm}

{\bf The Weak energy condition} is 
\be \label{weak}
\tilde{T}_{\mu\nu}W^{\mu}W^{\nu}\geq 0\, ,
\ee
where $W^{\mu}$ is a timelike vector.

\vspace{1cm}
The effective energy-momentum tensor can be rewritten as
\be \label{Tdec}
\tilde{T}_{\mu\nu}=\tilde{\rho} W_{\mu}W_{\nu}+\tilde{p}[g_{\mu\nu}+W_{\mu}W_{\nu}+\tilde{\Pi}_{\mu\nu}+2W_{(\mu}\tilde{q}_{\nu)}]\, ,
\ee
where 
\be \label{q1}
\tilde{\rho}=\tilde{T}_{\mu\nu} W^{\mu}W^{\nu},\,\,\,
\tilde{p}=\frac{1}{3}\tilde{T}_{\mu\nu}h^{\mu\nu}, \,\,\,
\tilde{q}^{\mu}=W^{\rho}\tilde{T}_{\rho \nu} h^{\mu\nu}\, ,
\ee
\be \label{q4}
\tilde{\Pi}^{\mu\nu}=\left(h^{\mu\rho}h^{\nu\sigma}-\frac{1}{3}h^{\mu\nu}h^{\rho\sigma} \right)\tilde{T}_{\rho\sigma},\,\,\,\,\,h^{\mu\nu}=g^{\mu\nu}+W^{\mu}W^{\nu}\, ,
\ee
where $\tilde{p},\tilde{\rho}$ are effective pressure and energy-density, 
$\tilde{q}^{\mu}$ is the effective energy/heat flow vector,
$\tilde{\Pi}^{\mu\nu}$ is the effective anisotropic stress tensor,
$h^{\mu\nu}$ is the induced metric on the Chauchy hypersurface. 

\section{Antievaporation and energy conditions}

In this section, we will discuss a generic Nariai antievaporating solution
and its relations with energy conditions. 
The Narai space-time is 
\be \label{Narai}
ds^{2}=\frac{1}{\Lambda}\left[-\frac{1}{\cosh^{2} x}(dx^{2}-dt^{2})+d\Omega^{2}_{D} \right]
\ee
where $\Lambda$ is the cosmological constant of De Sitter, 
$d\Omega^{2}_{D}$ is the solid angle on a $(D-2)$-sphere
and the scalar curvature is constant $R\sim \Lambda$. 
The stability of this solution can be 
with methods of the perturbation theory. 
A convenient parametrization can be 
\be \label{Narai2}
ds^{2}=e^{-\rho(x,t)}(dx^{2}-dt^{2})+e^{-\phi(x,t)}d\Omega^{2}_{D} 
\ee
where 
\be \label{rho}
\rho(x,t)=-{\rm ln}(\sqrt{\Lambda}\cosh x)+ \delta \rho(x,t)
\ee
\be \label{phi}
\phi(x,t)= {\rm ln} \sqrt{\Lambda}+\delta \phi(x,t)
\ee
If $\delta \phi(x,t)=-f(x,t)$
where $f(x,t)$
 is a growing function in time,
the Nariai solution will have an antievaporation instability
because of the Schwarzschild radius will grow as 
\be \label{grow}
r_{BH}\sim \frac{1}{\sqrt{\Lambda}}e^{f(t)}
\ee

As a consequence, that external time-like surfaces will be attracted inside the BH interior. 
As shown in \cite{Ellis:2013oka}, 
 an emitting marginally outer 2-surface $\mathcal{A}_{time-like}$
and the non-emitting one can be defined from the optical Raychaudhuri equation. The divergence of the outgoing null geodesics 
is defined as
$\hat{\theta}_{+}$ in a $S^{2}$-surface.

$\hat{\theta}_{+}$ decreases with the increase of the gravitational field. 
The divergence
of ingoing null geodesics is $\hat{\theta}_{-}<0$ 
in the entire black hole metric.
  $\hat{\theta}_{+}>0$ for $r>2m$ in Schwarzschild, while it is negative in the BH interior.  
The marginally outer trapped 2-surface $\mathcal{A}^{2d}_{MOT}$
is defined as a space-like 2-sphere satisfying the condition
\be \label{theta}
\hat{\theta}_{+}(\mathcal{A}_{MOT}^{2d})=0
\ee
In a Schwarzschild Black hole,
the radius of the sphere $\mathcal{A}_{MOT}^{2d}$
coincides with the Schwarzschild radius;
while $S^{2}$-spheres radii  
$r<r_{S}=2M$ are trapped surfaces, i.e. 
$\theta(\mathcal{A}^{2d}_{TS})<0$.

By virtue of definitions in 2d,
one can iteratively construct definitions for d-dimensional surfaces.
 In $D=4$, the surfaces of interest have 
$D-1=3$ dimensions. 
For an antievaporating BH in $D=4$, the horizon is a dynamical Marginally outer trapped 3-surface,
while in $D$-dimension a $D-1$-surface.  
The $D-1$ surface can be sectioned in a foliation of marginally trapped $D-2$-surfaces,
in turn foliated in $D-3$ and iteratively down to 2-surfaces. 
For simplicity, we can foliate the $D-1$-hypersurface in hyperspheres as the following chain
$$A^{D-2}_{TS}\rightarrow \mathcal{F}_{D-2}\{S^{D-3}\}\rightarrow ... \rightarrow \mathcal{F}_{1}\{S^{2}\}$$
where $\mathcal{F}_{D-2}$ are operations of foliation of a $D-2$ surface.

The difference between an emitting marginally outer trapped 
 $D-1$-surface 
$\mathcal{A}_{time-like}^{D-1}$
and 
a non-emitting surface 
$\mathcal{A}_{space-like}^{D-1}$
is exactly characterized by their different
directional derivative of $\hat{\theta}_{m}$
 along
 an ingoing null tangent vector $n_{a}$:
\be \label{fds}
\hat{\theta}_{m}(\mathcal{A}^{D-1}_{time-like})=0,\,\,\,\,\,\,\nabla_{n^{a}} \hat{\theta}_{m}(\mathcal{A}^{D-1}_{time-like})>0
\ee
while the non-emitting one is define as 
\be \label{sds}
\hat\theta_{m}(\mathcal{A}^{D-1}_{space-like})=0,\,\,\,\,\,\,\nabla_{n^{a}} \hat{\theta}_{m}(\mathcal{A}^{D-1}_{space-like})<0\, .
\ee

We can consider the Raychaudhuri equation associated to an antievaporating Nariai BH.
Let us solve the problem with choosing an initial Chauchy's condition $\theta(\bar{\lambda})>0$,
where $\bar{\lambda}$ is an initial condition of a generic affine parameter
$\lambda$. Generically for antievaporation, 
the null Raychauduri equation is bounded (setting $8\pi G_{N}=1$)
\be \label{boundR}
\frac{d\hat{\theta}}{d\lambda}<-(\mathcal{G}^{-1}T_{ab}+H_{ab})k^{a}k^{b}\, ,
\ee
with $T_{ab}k^{a}k^{b}=0$ on the BH horizon, 
the Cauchy surface can be focalized 
by $H_{ab}$ if
$$H_{ab}k^{a}k^{b}>0\, .$$

This leads us to formulate a general criteria for antievaporation: 
the extension of GR considered has to provide a
 an extra attractive term in the Raychauduri equation.
 Otherwise antievaporation cannot be sourced. 
 If and only if this condition is realize at the Schwarzschild radius of the Nariai solution, 
 the antievaporation phenomena will happen. 
In other words, even if no matter-energy was present in close to the BH horizon,
{\it the geometric fluid would source the focalization of the null Chauchy surface}. 
For this motivation, the Narai solution cannot be stable 
if $Hkk>0$ (tensor contraction omitted).  

In general, the combination
$Hkk$
will be a function of time with respect to an external observer in an inertial reference frame,
i.e.  $Hkk=H(x,t)$.
Now, if $H(x,t)$ is analytic function with respect to the time variable, 
we can perform a power series expansion. 
In this case, for antievaporating solution it will be always true that 
$Hkk>H_{0}\geq 0$
 where $H_{0}$ is the zero order coefficient of the Taylor expansion of $H(x,t)$ in time.  
This implies that 
\be \label{rett}
\hat{\theta}(\lambda)<\hat{\theta}(\lambda)-H_{0}(\lambda-\bar{\lambda})+O\{(\lambda-\bar{\lambda})^{2}\}
\ee
in turn implying 
$\hat{\theta}(\lambda)<0$ for $\lambda>\lambda_{1}-\hat{\theta}_{1}/H_{0}$ -- neglecting $O\{(\lambda-\bar{\lambda})^{2}\}$.

So that, we arrive to a general and powerful conclusion: 
{\it the antievaporation phenomena cannot happen if 
energy conditions, and in particular the positive null energy condition, 
for the tensor
$H_{ab}$
is violated!}.
In other words, {\it the positive null energy condition has to be respected 
not only by ordinary matter but also by the new geometrical fluid. }
In this sense, the null energy condition is deformed and it is extended to the 
effective energy-momentum tensor including the contribution of the geometrical fluid. 
This conclusion is connected to the attractive nature of gravity: a violation of the null 
energy condition of the entire effective energy-momentum tensor implies repulsiveness of gravity.   

\subsection{The case of $f(R)$-gravity}

In this section, we specialize our general arguments to the case of $f(R)$-gravity,
where
\be \label{HabfR}
\mathcal{G}_{1}=f'(R),\,\,\,\mathcal{G}_{2}=1,\,\,\,\,\,\,H_{\mu\nu}=\frac{1}{f'(R)}\left\{\frac{1}{2}[Rf'(R)-f(R)]g_{\mu\nu}-\nabla_{\mu}\nabla_{\nu}f'(R)+g_{\mu\nu}\Box f'(R) \right\}
\ee

The antievaporation condition is 
\be \label{anti}
H_{\mu\nu}k^{\mu}k^{\nu}=\left\{\frac{1}{2}[Rf'(R)-f(R)]g_{\mu\nu}k^{\mu}k^{\nu}-k^{\mu}k^{\nu}\nabla_{\mu}\nabla_{\nu}f'(R)+k^{\mu}k^{\nu}g_{\mu\nu}\Box f'(R) \right\}>0\,.
\ee

Let us consider this condition in 
the Nariai background. 
Now we can use the perturbation theory methods around the Nariai solution, using the parametrization of Eq.(\ref{Narai2}), 
we obtain from Eq.(\ref{anti}).
Since, we are interested to study fluctuations 
on the horizon,
where $G_{\mu\nu}k^{\mu}k^{\nu}=0$, 
we can perturbe 
$(G_{\mu\nu}+H_{\mu\nu})k^{\mu}k^{\nu}$ 
equivalently to $H_{\mu\nu}k^{\mu}k^{\nu}$.
We can decompose Eq.(\ref{anti}) in components $(t,t),(x,x),(t,x), (\theta,\theta), (\psi,\psi)$
-- conditions for $(\theta,\theta)$ and $(\psi,\psi)$ lead to the same equation --
equivalent to the following equations. We will try the extremal solutions for $\delta(G_{\mu\nu}+H_{\mu\nu})k^{\mu}k^{\nu}=0$:
\be \label{eq00}
0=\frac{-f'(R_{0})+2\Lambda f''(R_{0})}{2\Lambda \cosh^{2}x}[4\Lambda (-\delta \rho+\delta \phi)+\Lambda \cosh^{2}x(2\delta \ddot{\rho}-2\delta \rho'-4\delta \ddot{\phi}+\delta \phi'')]
\ee
$$-\frac{f(R_{0})}{\Lambda \cosh^{2}x}\delta \rho-f'(R_{0})(-\delta \ddot{\rho}+2\delta \ddot{\phi}+\delta \rho''+2\tanh x \delta \phi')+(\delta R''+\tanh x \delta R') f''(R_{0})=0$$

\be \label{eq11}
0=\frac{-f'(R_{0}+2\Lambda^{2}f''(R_{0})}{2\Lambda \cosh^{2}x}[4\Lambda (-\delta \rho+\delta \phi)+\Lambda \cosh^{2}x(2\delta \ddot{\rho}-2\delta \rho'-4\delta \ddot{\phi}+\delta \phi'')]
\ee
$$+	\frac{f(R_{0})}{\Lambda \cosh^{2}x}\delta \rho-f'(R_{0})(\delta \ddot{\rho}+2\delta \phi''-\delta \rho''+2\tanh x \delta \phi')
+f''(R_{0})\delta \ddot{R} $$

\be \label{eq01}
0=-2(\delta \dot{\phi}'+\tanh x \delta \dot{\phi})+\frac{f''(R_{0})}{f'(R_{0})}(\delta \dot{R}'+\tanh x \delta \dot{R})
\ee
\be \label{eqthetatheta}
0=[4\Lambda (-\delta \rho+\delta \phi)+\Lambda \cosh^{2}x(2\delta \ddot{\rho}-2\delta \rho'-4\delta \ddot{\phi}+\delta \phi'')]\frac{-f'(R_{0})+2\Lambda f''(R_{0})}{2\Lambda }
\ee
$$-\frac{f(R_{0})}{\Lambda }\delta \phi-\cosh^{2}x f'(R_{0})(-\delta \ddot{\phi}+\delta \phi'')-\cosh^{2}(-\delta \ddot{R} +\delta R'')f''(R_{0})$$
where 
$$\delta R=\Lambda [\cosh^{2}x(2\delta \ddot{\rho}-2\delta \rho''-4\delta \ddot{\phi}+4\phi'')-4\delta \rho+4\delta \phi)]$$

These equations correspond to the ones studied by Nojiri and Odintsov.
They obtained the same set of equations starting form field equations while 
in our case from the antievaporation bound \cite{Nojiri:2013su}. 
As already tried in Ref. \cite{Nojiri:2013su},
analyzing these equations, one can obtain the 
solution 
\be \label{deltaphiphi}
r_{H}^{-2}=\frac{1+\delta \phi(x_{0},\tau)}{\Lambda},\,\,\,
\delta \phi_{\pm}=\phi_{0}\,{\rm exp}\left\{\frac{-1\pm \sqrt{1+4\mu^{2}}}{2}t\right\},\,\,\, \mu^{2}=\frac{2(2\alpha-1)}{3\alpha},\,\,\,\alpha=\frac{2\Lambda f_{RR}(R_{0})}{f_{R}(R_{0})}\, .
\ee

Considering $\delta \phi$, 
it will antievaporate if 
$\mu^{2}>0$, i.e.
$\alpha<0\, \&\, \alpha>1/2$. 
This bound is obtained from condition in Eq.(\ref{anti})
On the other hand, for $4\mu^{2}<0$
the solution evaporates, corresponding to the opposite bound to Eq.(\ref{anti}). 

\subsection{The case of Gauss-Bonnet gravity}
In this section, 
we will consider the Nariai black holes in $f(G)$-gravity. 
The $f(G)$-gravity has an action which reads 
\be \label{action}
I=\frac{1}{2\kappa^{2}}\int d^{4}x\sqrt{-g}[R+f(G)]
\ee
where $G$ is the Gauss-Bonnet topological invariant
\be \label{GB}
G=R^{2}-4R_{\mu\nu}R^{\mu\nu}+R_{\mu\nu\rho\sigma}R^{\mu\nu\rho\sigma}\, .
\ee
The EoM in vacuum is 
\be \label{RRR}
G_{\mu\nu}=-H_{\mu\nu}
\ee
where 
\be \label{Hmunu}
-H_{\mu\nu}=\frac{1}{2}g_{\mu\nu}f-2FRR_{\mu\nu}+4FR_{\mu\rho}R^{\rho}_{\nu}
\ee
$$-2FR_{\mu}^{\rho\sigma \tau}R_{\nu\rho\sigma \tau}-4FR_{\mu\nu}^{\rho\sigma}R_{\rho\sigma}
+2R\nabla_{\mu}\nabla_{\nu}F-2Rg_{\mu\nu}\nabla^{2}F$$
$$-4R_{\nu}^{\rho}\nabla_{\rho}\nabla_{\mu}F-4R_{\mu}^{\rho}\nabla_{\rho}\nabla_{\nu}F
+4R_{\mu\nu}\nabla^{2}F+4g_{\mu\nu}R^{\rho\sigma}\nabla_{\rho}\nabla_{\sigma}F
-4R_{\mu\nu}^{\rho\sigma}\nabla_{\rho}\nabla_{\sigma}F\, . $$
Here we used the notation $F=\partial f/\partial G$. 

The antievaporation condition, related to the positive null energy condition 
on the H tensor, must be applied on Eq.(\ref{Hmunu}). 
Perturbing the antievaporation condition as $\delta (G+H)kk|_{horizon}$, 
which is equivalent to $\delta Hkk$ on the black hole horizon, 
we obtain
\be \label{EoMG1}
4\Lambda(F(G_{0})+\Lambda^{2}\cos^{2}\tau)\delta \ddot{\rho}
+\left( 2+8\Lambda(F(G_{0}\sec^{2}\tau-\Lambda \cos^{2}\tau)\right)\delta \rho''
\ee
$$-2\tan \tau\left(1+4\Lambda F(G_{0})-4\Lambda (F(G_{0})\sec^{2}\tau+\Lambda\cos^{2}\tau)\right)\delta \dot{\phi}+\sec^{2}\tau\left(\frac{1}{2\Lambda}-2F(G_{0}) \right)\delta R$$
$$+\delta \rho\left( \sec^{2}\tau(4+\Lambda^{-1}f(G_{0})-12\Lambda F(G_{0}))-8\Lambda^{2}\right)+\sec^{2}\tau\left(\frac{F(G_{0})}{2\Lambda}-2\Lambda F'(G_{0}) \right)\delta G+8\Lambda F'(G_{0})\delta \dot{G}'$$
$$+4\Lambda F'(G_{0})\delta \ddot{G}-8\Lambda \tan \tau F'(G_{0})\delta \dot{G}+4\Lambda F'(G_{0})\delta G''-8\Lambda \tan \tau F'(G_{0})\delta G'+8\Lambda^{2}(\delta \phi-\delta \rho)=0\, ,$$

\be \label{EoMG2}
(-1+4\Lambda F(G_{0}))\delta \rho''+(1+4\Lambda F(G_{0}))\delta \ddot{\rho}+2\delta \phi''
-2\tan \tau (1-4\Lambda F(G_{0}))\delta \dot{\phi}
\ee
$$+\sec^{2}\tau\left( -4-\Lambda^{-1}f(G_{0})+4\Lambda F(G_{0})\right)\delta \rho+\sec^{2}\tau\left( 2F(G_{0})-\Lambda^{-1}\right)\delta \dot{\phi}$$
$$+\sec^{2}\tau\left( -4-\Lambda^{-1}f(G_{0})+4\Lambda F(G_{0})\right)\delta \rho+\sec^{2}\tau\left(2F(G_{0})-\frac{1}{2\Lambda} \right)\delta R$$
$$+\sec^{2}\tau\left(-\frac{F(G_{0})}{2\Lambda}+6\Lambda F'(G_{0})\right)\delta G+4\Lambda F'(G_{0})(\delta G''-\delta \ddot{G})=0\, , $$

\be \label{EoMG3}
\left(1+4\Lambda F(G_{0})(3-2\cos^{2}\tau\right)\delta \dot{\phi}'-\tan \tau(1+8\Lambda F(G_{0})\sin^{2}\tau)\delta \phi'-6\Lambda
F'(G_{0})(\delta \dot{G}'-\tan \tau \delta G')=0\, ,
\ee

\be \label{EoMG4}
-\cos \tau \left(\cos \tau+4\Lambda F(G_{0})(1-\cos^{3}\tau) \right)\delta \ddot{\phi}
+\cos \tau \left( \cos \tau+4\Lambda F(G_{0})\right)\delta \phi''
\ee
$$+\left(2F(G_{0})-\frac{1}{2\Lambda}-\frac{4F(G_{0})}{\Lambda} \right)\left(\frac{4}{\Lambda}+\frac{f(G_{0})}{\Lambda}+\frac{4\Lambda F(G_{0})}{\cos \tau}\right) \delta R\delta \phi+\left(8\Lambda F'(G)-\frac{4F'(G_{0})}{\Lambda}
-2\Lambda F'(G_{0})\sec^{2}\tau-\frac{F(G_{0})}{2\Lambda}\right)\delta G   $$
$$+4\Lambda F(G_{0})\sec \tau(1-\sec \tau)\delta \rho-4\Lambda F(G_{0})\cos^{4}\tau \tan \tau \delta \dot{\phi}=0\, .$$
Here we defined  the perturbation around the Gauss-Bonnet invariant and Ricci scalar backgrounds $G_{0}$ and $R_{0}$
as follows: 
\be \label{deltaG}
\delta G=G_{0}\left[ \delta \ddot{\rho}-\frac{5}{2}\cos^{2}\tau \delta \rho''+2\cos^{2}\tau \delta \phi+2(\cos^{2}\tau-2)\delta \rho\right]
\ee
\be \label{deltaR}
\delta R=2F'(G_{0})\left(G_{0}\delta G+R_{0}\box \delta G-2R_{\mu\nu}^{0}\nabla^{\mu}\nabla^{\nu}\delta G \right)\, . 
\ee

As tried in $f(R)$-case studied above, the system of equations that we obtain is exactly the one studied in Ref. \cite{Sebastiani:2013fsa}
and leading to antievaporating solutions. 
We limit our-self to comment on the case studied numerically in Ref. \cite{Sebastiani:2013fsa} and displayed in Fig.10. 
The authors considered a model for accelerated cosmology
\be \label{fGl}
f(G)=\lambda\sqrt{G_{S}}\left[-\alpha+g(x)\right],\,\,\,x=G/G_{S},\,\,\, g(x)=x\arctan x-\frac{1}{2}{\rm log}(1+x^{2})\, , 
\ee
where $\alpha,\lambda$ are real positive parameters and $G_{s}\sim H_{0}^{4}$ with $H_{0}$ the Hubble constant. 
The numerical solution shown in Fig.10 of Ref. \cite{Sebastiani:2013fsa}
shown a secular antievaporation which is followed later on by a violent evaporating 
contraction. This means that energy conditions have dynamically switched from 
positive NEC preserving $Hkk>0$ to the NEC violating $Hkk<0$. 

\section{Conclusions and outlooks}

In this paper, we have studied evaporation/antievaporation in extended theories of gravity. 
From our analysis, we arrived to the conclusion that evaporating/antievaporating Nariai solution will be 
omnipresent
 in Lorentz invariant extensions of the Einstein-Hilbert action
\footnote{The problem of classical stable BH solution 
in Lorentz Breaking massive gravity was studied in our paper
\cite{Addazi:2014mga}. We found stringent constraints from the 
analysis of geodetic stability.}. 
In particular, we related the evaporation to the violation
of the null energy condition 
imposed on the geometric fluid -- composed of extra degree of freedom coupled to the massless 
spin-2 graviton. On the contrary, a violation of the null energy condition (NEC) by the geometric fluid
turns off the antievaporation instability. 
Our results are independent by the number of space-time dimension, so that 
they can be applied for example for solutions of string-inspired modified gravities
like brane-worlds or intersecting D-brane worlds 
embedded in a higher dimensional bulk. 

So that, the imposition of the NEC on the gravitational energy-momentum tensor 
necessary {\it censors} any possible evaporation instabilities. 
On the other hand, if the NEC condition bound is
positive, antievaporation inevitably will occur. 
In order to turn off any instabiltiies, the 
gravitational energy-momentum tensor must be 
null on the black hole horizon.

Finally, let us comment that still many aspects on 
evaporation and antievaporation remain obscure. 
First of all, it is unclear what this should imply at virtual level,
in particular in the case 
 of virtual black holes.
 This issue is very relevant for issues regarding the cosmological 
 vacuum energy \cite{Addazi:2016jfq} and information in quantum gravity \cite{Addazi:2017xur}.
Second, further quantum chaos effects are expected to be
crucially important in such an unstable solutions 
  \cite{Addazi:2015cho,Addazi:2016cad,Addazi:2015gna}.
  Finally, it is conjecturable that further other reasons 
related to new symmetry principles 
  could be found behind evaporation/antivaporation 
  dynamics -- see e.g. Refs.\cite{Paliathanasis:2011jq,Paliathanasis:2014rja,Paliathanasis:2015aos}.

\begin{acknowledgments} 

I would like to thank Salvatore Capozziello, Antonino Marciano and Sergei Odintsov for valuable comments and discussions
on these subjects.

\end{acknowledgments}

\end{document}